\documentclass[dvips]{article}
\usepackage{gea}
\usepackage{icrctc07}

\title{Nonthermal Bremsstrahlung vs. Synchrotron Radiation: Cas A}
\shorttitle{Nonthermal Bremsstrahlung vs. Synchrotron Radiation: Cas A}
\authors{G.\ E.\ Allen$^{1}$, M.\ D.\ Stage$^{2}$, J.\ C.\ Houck$^{1}$.}
\shortauthors{G.\ E.\ Allen, et~al.}
\afiliations{$^{1}$Massachusetts Institute of Technology, Kavli Institute
for Astrophysics and Space Research, 77 Massachusetts Avenue, NE80,
Cambridge, MA 02139-4307\\ $^{2}$University of Massachusetts, Department of
Astronomy, LGRT-B 619E, 710 North Pleasant Street, Amherst, MA 01003-9305}
\email{gea@space.mit.edu}

\abstract{We performed a spectral analysis of some {\sl Chandra} ACIS and
{\sl RXTE} PCA data for the supernova remnant Cas A.  A very large (1.1~Ms)
ACIS data set is used to identify regions dominated by synchrotron
radiation.  The best-fit spectral models for these regions are combined to
obtain a composite synchrotron model for the entire remnant.  The difference
between this model and the observed PCA flux is fitted with a nonthermal
bremsstrahlung model. The results of this analysis suggest that (1) the
ratio of the nonthermal bremsstrahlung to synchrotron radiation varies from
about 2:1 to 4:1 in the 10--32~keV energy band, (2) the electron spectrum is
significantly steeper at 10--32~keV than it is at 1~GeV, (3) about 5\% of
the electrons are nonthermal and (4) about 30\% of the energy in the
electron distribution is in nonthermal electrons.}

\begin{document}
\maketitle

\section{Introduction}

The nature of the high-energy X-ray emission from Cas~A has been a source of
controversy for at least the last decade. Some have suggested that the
emission is dominated by synchrotron radiation from TeV electrons
accelerated at the forward shock of the remnant \cite{all97a}.  Others argue
that most of the emission is produced by nonthermal bremsstrahlung emission
from electrons that have energies only slightly larger that the thermal
electron energies \cite{ble01a,lam01a, vin03a}. The goal of the present
analysis is to determine how much each mechanism contributes to the
high-energy X-ray spectrum.  The results will help determine the properties
of the nonthermal electrons at energies just above the thermal peak.

\section{Data and Analysis}

\begin{figure}
  \begin{center}
    \includegraphics [width=0.48\textwidth]{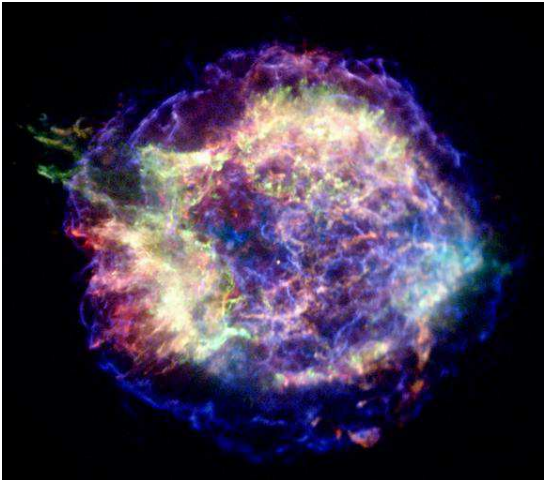}
  \end{center}
  \caption{A 1.1~Ms, color-coded, ACIS image of the supernova remnant Cas~A.
    The red, green and blue features are associated with the 0.5--1.5 keV
    (O, Fe~L, Ne and Mg emission-line), 1.5--2.5 (Si and S emission-line)
    and 4--6 keV (continuum) energy bands, respectively.}
  \label{fig01}
\end{figure}

\begin{figure}
  \begin{center}
    \includegraphics [width=0.48\textwidth]{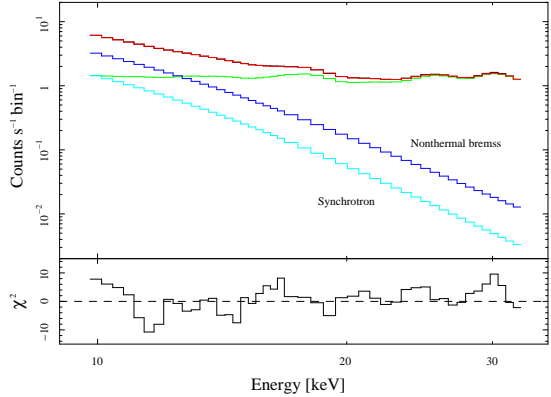}
  \end{center}
  \caption{The PCA data (black) and background (green) spectra, the global
    synchrotron spectrum (light blue) and the nonthermal bremsstrahlung
    spectrum (dark blue) are plotted in the top panel. The red histogram is
    the sum of the background, synchrotron and nonthermal bremsstrahlung
    spectra. The bottom panel shows the differences between the nearly
    identical black and red histograms divided by the square root of the
    black histogram.}
  \label{fig02}
\end{figure}

\begin{figure}
  \begin{center}
    \includegraphics [width=0.48\textwidth]{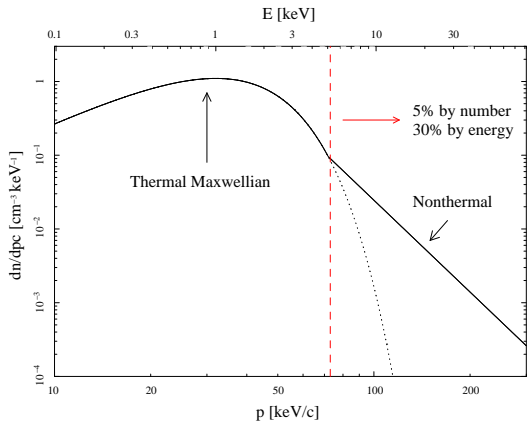}
  \end{center}
  \caption{The thermal and nonthermal electron spectrum for the entire
    supernova remnant Cas~A.  The dashed red line marks the transition
    between the thermal and nothermal bands.  The dotted black curve shows
    what the thermal Maxwellian distribution would be if the remnant did not
    contain nonthermal electrons.}
  \label{fig03}
\end{figure}

The present work is based on an analysis of 1.1~Ms of {\sl Chandra} ACIS
data \cite{hwa04a, sta06a} and 91~ks of {\sl RXTE} PCA data \cite{all97a}.
As described by \cite{sta06a}, the 0.3--7~keV ACIS data are used to identify
the regions dominated by synchrotron radiation. These regions are the ones
with a relatively large fraction of their emission in the 4--6~keV band
(\ie\ the blue filaments in Fig.~\ref{fig01}) or, equivalently, the ones
with relatively high apparent electron temperatures (Fig.~2 of
\cite{sta06a}).
Using a technique similar to the one described by \cite{sta06a}, the
0.3--7~keV ACIS spectrum for each one of 1,896 nonoverlapping regions
dominated by synchrotron radiation was fitted with a model that includes an
interstellar absorption component and a synchrotron component. The composite
0.3--7~keV synchrotron spectrum of the entire remnant is obtained by summing
the best-fit spectral models for each of the synchrotron-dominated regions.
This composite model is extrapolated to higher X-ray energies and compared
to the 10--32~keV PCA spectrum for Cas~A\footnote{Note that the PCA is not
an imaging instrument.  It is not possible to separately measure the spectra
of small regions of the source. Therefore, the spectra of the small regions
studied using the ACIS data must be summed before they are compared to the
PCA data.}. The PCA data below 10~keV are excluded, to simplify the
analysis. A thermal emission component and an interstellar absorption
component can be neglected at energies greater than 10~keV. The data above
32~keV are ignored because the signal-to-noise ratio is less than one in
this range. When the composite synchrotron spectrum derived from the ACIS
data is compared to the PCA data, the results indicate that only a minority
of the PCA flux is produced by synchrotron radiation. Therefore the PCA data
were fitted with a model that includes two components: a synchrotron
component and a nonthermal bremsstrahlung component.  These components are
described in detail by \cite{hou06a}.  The synchrotron component is the
composite synchrotron spectrum as extrapolated to PCA energy band.  This
component has no free parameters in the fit. Therefore, we assume that the
relative calibration uncertainties of the ACIS and the PCA are negligible.
The 10--32~keV emission in excess of the synchrotron radiation is fitted
with a nonthermal bremsstrahlung component.  While this model includes the
low-energy correction of \cite{elw39a}, the model is based on the assumption
that the target particles are at rest.  The nonthermal bremsstrahlung
component has two free parameters: the differential spectral index of the
nonthermal electron spectrum ($\Gamma$) and the normalization of the
nonthermal bremsstrahlung emission. The best-fit parameters are listed in
Table~1.
\begin{table}
  \begin{center}
    \begin{tabular}{l c}
      \multicolumn{2}{c}{Table 1. Best-fit parameters}
        \\
      \hline
      \hline
      \multicolumn{1}{c}{Parameter}
        & Value
        \\
      \hline
      $\Gamma$
        & 4.14
        \\
      Norm [cm$^{-5}$ at 1~GeV]
        & 11.6
        \\
      \hline
      \hline
    \end{tabular}
  \end{center}
\end{table}
The data (black histogram) and model (red histogram) for the PCA data are
plotted in Figure~\ref{fig02}.  Most of the black histogram is obscured by
the red histogram because these two histograms are nearly identical.  The
red histogram is the sum of the green (instrumental background), light blue
(synchrotron model) and dark blue (nonthermal bremsstrahlung model)
components.  As shown, the energy-dependent ratio of the nonthermal
bremsstrahlung and synchrotron components is about 2:1 to 4:1 in the
10--32~keV band.
A combination of the results for the nonthermal bremsstrahlung model and the
results for a separate, thermal bremsstrahlung analysis ($kT = 1$~keV, ${\rm
Norm} = 2.32$~cm$^{-5}$) are used to infer the properties of the global
electron spectrum. This spectrum, which is shown in Figure~\ref{fig03},
includes the following features:
\begin{itemize}
  \item
    There is a transition from the thermal Maxwellian to a nonthermal
    distribution at $E = 5.3$~keV ($p = 73$~keV/$c$).
  \item
    About 5\% of the electrons have energies greater than 5.3~keV.  This is
    one measure of the efficiency with which electrons are injected into the
    acceleration process in the remnant.
  \item
    The nonthermal electrons contain about 30\% (\ie\ 2 $\times$
    10$^{49}$~ergs) of the energy in the entire electron distribution.
\end{itemize}

\section{Conclusions}

By expanding on a previous analysis of the {\sl Chandra} ACIS data for the
supernova remnant Cas~A \cite{sta06a}, we have tried to separate the
synchrotron emission of the remnant from the bremsstrahlung emission in the
ACIS energy band (0.3--7~keV).  An extrapolation of the total synchrotron
model for the remnant from the ACIS energy band to the 10--32~keV energy
band enabled us to model the excess PCA emission using a nonthermal
bremsstrahlung component.  The results of this analysis indicate that:
\begin{itemize}
  \item
    About 70--80\% of the 10--32~keV emission is produced by nonthermal
    bremsstrahlung (Fig.~\ref{fig02}). The balance of the emission is
    produced by synchrotron radiation.
  \item
    The differential spectral index of the electrons that produce the
    nonthermal bremsstrahlung emission ($\Gamma$ = 4.1) is significantly
    larger than the index of the radio-synchrotron--producing electrons
    ($\Gamma$ = 2.54, \cite{baa77a}).  This result may indicate that the
    spectral slope of the injected electrons is 4.1.
  \item
    A substantial fraction of the energy in the electron distribution is in
    nonthermal electrons (30\%).  If the same is true for nonthermal
    protons, then this result supports the idea that the shock has been
    modified by cosmic rays (\ie\ cosmic rays are not test particles).
\end{itemize}

These conclusions should be regarded as preliminary because a number of
potential concerns have not yet been investigated.  These issues include:
\begin{itemize}
  \item
    the possibility that the relative calibration of the {\sl Chandra} ACIS
    and {\sl RXTE} PCA is not negligible.  Calibration differences could
    affect the fitted index and normalization of the nonthermal
    bremsstrahlung model and, hence, the properties of the inferred electron
    distribution.
  \item
    the possibility that the composite synchrotron spectrum is inaccurate
    because some of the X-ray synchrotron emission in the ACIS data has been
    overlooked.  If this is true, then the amount of emission attributed to
    nonthermal bremsstrahlung may be too high.
  \item
    the assumption that the target particles in bremsstrahlung interactions
    are at rest.  The cross section used for the nonthermal bremsstrahlung
    model may have to be modified.
\end{itemize}

\section{Acknowledgements}

GEA and JCH are supported by the contract SV3-73016 between MIT and the
Smithsonian Astrophysical Observatory.  The Chandra X-Ray Center at the
Smithsonian Astrophysical Observatory is operated on behalf of NASA under
the contract NAS8-03060. MDS has been supported by the NASA LTSA grant
NAG5-9237 and the Five College Astronomy Department Fellowship program.
\bibliography{icrc1175}
\bibliographystyle{plain}

\end{document}